\documentclass{ws-ijmpd}
\usepackage[super,compress]{cite}
\begin{document}

\markboth{Remo Garattini}
{Effects of additional sources on Casimir Wormholes}

%
\catchline{}{}{}{}{}
%

\title{Effects of additional sources on Casimir Wormholes}

\author{Remo Garattini}
\address{Universit\`{a} degli Studi di Bergamo,\\ Dipartimento di Ingegneria e Scienze
Applicate,\\Viale Marconi 5, 24044 Dalmine (Bergamo) Italy and\\
I.N.F.N. - sezione di Milano, Milan, Italy.\\
remo.garattini@unibg.it}

\maketitle

\begin{history}
\received{Day Month Year}
\revised{Day Month Year}
\end{history}

\begin{abstract}
In this contribution we explore the consequences of including additional sources to the original Casimir energy Stress-Energy Tensor. In particular, we will discuss the effects of an additional electromagnetic field, the modification induced by non-zero temperature effects on the energy density obtained by a Casimir device and finally the effect obtained by including a massless scalar field. For each of these examples, we have introduced an auxiliary stress tensor which we have interpreted as a thermal tensor. Consequences on the size of the throat are also discussed. We will show that these additional extra fields do not destroy the traversability of the wormhole.
\end{abstract}

\keywords{Traversable Wormholes; Casimir effect.}

\ccode{PACS numbers:}

\tableofcontents

\section{Introduction}	

In Quantum Field Theory, the Zero Point Energy (ZPE) represents a
fundamental physical quantity that must be computed to understand the
properties of the physical system under examination. A particular ZPE case
is the Casimir Energy. This is obtained when ons studies quantum
fluctuations between two plane parallel, closely spaced, uncharged, metallic
plates in vacuum. It was predicted theoretically in 1948\cite{Casimir} and
experimentally confirmed in the Philips laboratories\cite{Sparnaay,AVS}.
However, only in recent years further reliable experimental investigations
have confirmed such a phenomenon\cite{Lamoreaux,BCOR}. The interesting
feature of this effect is that an attractive force appears which is
generated by negative energy. Indeed the attractive force arises because the
renormalized energy assumes the following form%
\begin{equation}
E^{\text{Ren}}\left( a\right) =-\frac{\hbar c\pi ^{2}S}{720a^{3}},
\end{equation}%
where $S$ is the surface of the plates and $a$ is the separation between
them. The force can be obtained with the computation of%
\begin{equation}
F\left( a\right) =-\frac{dE^{\text{Ren}}\left( a\right) }{da}=-3\frac{\hbar
c\pi ^{2}S}{720a^{4}}.  \label{F(a)}
\end{equation}%
From the force $\left( \ref{F(a)}\right) $ is also possible to compute the
pressure%
\begin{equation}
P\left( a\right) =\frac{F\left( a\right) }{S}=-3\frac{\hbar c\pi ^{2}}{%
720a^{4}}.  \label{P(a)}
\end{equation}%
It is immediate to recognize that the energy density is nothing but%
\begin{equation}
\rho _{C}\left( a\right) =-\frac{\hbar c\pi ^{2}}{720a^{4}}  \label{rhoC}
\end{equation}%
suggesting the existence of a relation between the pressure $P$ and the
energy density $\rho $ described by an Equation of State (EoS) of the form $%
P=\omega \rho $ with $\omega =3$. The nature of this effect is connected
with the ZPE of the Quantum Electrodynamics Vacuum distorted by the plates.
It is important to observe that this effect has a strong dependence on the
geometry of the boundaries. Indeed, Boyer\cite{Boyer} proofed the positivity
of the Casimir effect for a conducting spherical shell of radius $r$. The
same positivity has been proofed also in Ref.\refcite{BVW}, by means of heat
kernel and zeta regularization techniques. From Eqs.$\left( \ref{rhoC},\ref%
{P(a)}\right) $, we can write down the form of the Stress-Energy Tensor
(SET) which is%
\begin{equation}
T_{\mu \nu }=-\frac{\hbar c\pi ^{2}}{720a^{4}}diag\left( 1,3,-1-1\right) ,
\label{SET}
\end{equation}%
where we have also introduced the expression of the transverse pressure%
\begin{equation}
p_{t}\left( a\right) =\frac{\hbar c\pi ^{2}}{720a^{4}}.
\end{equation}%
The SET $\left( \ref{SET}\right) $ is traceless and divergenceless and will
be used as a prototype source for an amazing object predicted by General
Relativity: a Traversable Wormhole (TW). Traversable wormholes (TW) are
solutions of the Einstein's Field Equations (EFE) powered by classical
sources\cite{MT,Visser}. However, given the quantum nature of the Casimir
effect, the EFE must be replaced with the semiclassical EFE, namely%
\begin{equation}
G_{\mu \nu }=\kappa \left\langle T_{\mu \nu }\right\rangle ^{\text{Ren}%
}\qquad \kappa =\frac{8\pi G}{c^{4}},
\end{equation}%
where $\left\langle T_{\mu \nu }\right\rangle ^{\text{Ren}}$ describes the
renormalized quantum contribution of some matter fields: in this specific
case, the electromagnetic field. To further proceed, we introduce the
following spacetime metric%
\begin{equation}
ds^{2}=-e^{2\Phi \!\left( r\right) }\,dt^{2}+\frac{dr^{2}}{1-b(r)/r}%
+r^{2}\,(d\theta ^{2}+\sin ^{2}{\theta }\,d\varphi ^{2})\,,  \label{metric}
\end{equation}%
representing a spherically symmetric and static wormhole. Here, $\Phi
\!\left( r\right) $ and $b(r)$ are arbitrary functions of the radial
coordinate $r\in \left[ r_{0},+\infty \right) $, and they represent the
redshift function, and the shape function, respectively \cite{MT,Visser}. An
important property of a traversable wormhole is the flare-out condition,
given by $(b-b^{\prime }r)/b^{2}>0$. It must be satisfied along with the
condition $1-b(r)/r>0$. Furthermore, at the throat $b(r_{0})=r_{0}$ is
satisfied, and the condition $b^{\prime }(r_{0})<1$ is imposed to obtain
wormhole solutions. It is also fundamental that there are no horizons
present, which are identified as the surfaces with $e^{2\Phi \!\left(
r\right) }\rightarrow 0$, so that $\Phi \!\left( r\right) $ must be finite
everywhere. With the help of the line element $\left( \ref{metric}\right) $,
we can write the EFE in an orthonormal reference frame, leading to the
following set of equations%
\begin{equation}
\frac{b^{\prime }\left( r\right) }{r^{2}}=\kappa \rho \left( r\right) ,
\label{rho}
\end{equation}%
\begin{equation}
\frac{2}{r}\left( 1-\frac{b\left( r\right) }{r}\right) \Phi \!^{\prime
}\left( r\right) -\frac{b\left( r\right) }{r^{3}}=\kappa p_{r}\left(
r\right) ,  \label{pr}
\end{equation}%
\begin{equation}
\left( 1-\frac{b\left( r\right) }{r}\right) \left[ \Phi ^{\prime \prime
}\!\left( r\right) +\Phi \!^{\prime }\left( r\right) \left( \Phi ^{\prime
}\!\left( r\right) +\frac{1}{r}\right) \right] -\frac{b^{\prime }\left(
r\right) r-b\left( r\right) }{2r^{2}}\left( \Phi \!^{\prime }\left( r\right)
+\frac{1}{r}\right) =\kappa p_{t}(r), \label{pt}
\end{equation}%
in which $\rho \left( r\right) $ is the energy density, $p_{r}\left(
r\right) $ is the radial pressure, and $p_{t}\left( r\right) $ is the
lateral pressure. We can complete the EFE with the expression of the
conservation of the stress-energy tensor which can be written in the same
orthonormal reference frame 
\begin{equation}
p_{r}^{\prime }\left( r\right) =\frac{2}{r}\left( p_{t}\left( r\right)
-p_{r}\left( r\right) \right) -\left( \rho \left( r\right) +p_{r}\left(
r\right) \right) \Phi \!^{\prime }\left( r\right) .
\end{equation}%
As far as we know, the Casimir energy represents the only artificial source
of \textit{exotic matter}\ realizable in a laboratory\footnote{%
Actually, there exists also the possibility of taking under consideration a
squeezed vacuum. See for example Ref.\refcite{Hochberg}.}\textit{. }The rest of
the paper is organized as follows, in section \ref{p2}, we analyze the
structure of the SET of the sources under investigation, in section \ref{p3}
we will describe what is a Casimir Wormhole (CW), in section \ref{p4} we
will include an electromagnetic source to the original Casimir source, in
section \ref{p5} we will reexamine the structure of the Casimir source by
discussing several alternatives solution proposals, in section \ref{p6} we
will analyze the effect of a non vanishing temperature on the TW, in section %
\ref{p7} we will examine the effect of a massless scalar field minimally
couple with a Casimir source. We summarize and conclude in section \ref{p8}.
Units in which $\hbar =c=k=1$ are used throughout the paper and will be
reintroduced whenever it is necessary.

\section{Structure of the Stress-Energy Tensor}

\label{p2}Consider an anisotropic fluid with radial pressure $p_{r}$, transverse
pressure $p_{t},$ energy density $\rho $ and a thermal stress tensor $\tau $\cite{Hayward}
described by the Stress-Energy Tensor (SET)%
\begin{equation}
T_{\mu \nu }=\left( \rho +\tau _{\rho }\right) u_{\mu }u_{\nu }+\left(
p_{r}+\tau _{r}\right) n_{\mu }n_{\nu }+\left( p_{t}+\tau _{t}\right) \sigma
_{\mu \nu },  \label{Tmn}
\end{equation}%
where $u_{\mu }$ is the fluid four-velocity and $n_{\mu }$ is a unit
spacelike vector orthogonal to $u_{\mu }$, i.e., $n^{\mu }n_{\mu }=1$, $%
u^{\mu }u_{\mu }=-1$, $n^{\mu }u_{\mu }=0$. The thermal stress tensor has
been decomposed into an energy component $\tau _{\rho}$, a radial component $%
\tau _{r}$ and a transverse component $\tau _{t}$. Here,%
\begin{equation}
\sigma _{\mu \nu }=g_{\mu \nu }+u_{\mu }u_{\nu }-n_{\mu }n_{\nu }
\end{equation}%
is a projection operator onto a two-surface orthogonal to $u_{\mu }$ and $%
n_{\mu }$, i.e.,%
\begin{equation}
u_{\mu }\sigma ^{\mu \nu }\mathrm{v}_{\nu }=n_{\mu }\sigma ^{\mu \nu }%
\mathrm{v}_{\nu }=0\qquad \forall \mathrm{v}_{\nu }.  \label{orto}
\end{equation}%
$n_{\mu }$ can be represented as%
\begin{equation}
n_{\mu }=\sqrt{\frac{1}{1-b(r)/r}}\left( 0,1,0,0\right) .
\end{equation}%
With these informations, we can write the components of the SET $\left( \ref%
{SET}\right) $%
\begin{align}
T_{tt}& =\left( \rho +\tau _{\rho }\right) u_{t}u_{t},  \label{Ttt} \\
T_{rr}& =\left( p_{r}+\tau _{r}\right) n_{r}n_{r},  \label{Trr} \\
T_{{\theta \theta }}& =\left( p_{t}+\tau _{t}\right) \sigma _{{\theta \theta 
}},  \label{Tthth} \\
T_{\varphi \varphi }& =\left( p_{t}+\tau _{t}\right) \sigma _{\varphi
\varphi }.  \label{Tpp}
\end{align}%
Since we desire to adopt a physical source, we assume that the quadruple $%
\left( \rho ,p_{r},p_{t},p_{t}\right) $ be represented by the Casimir SET $%
\left( \ref{SET}\right) $, while the quadruple $\left( \tau _{\rho },\tau
_{r},\tau _{t},\tau _{t}\right) $ has to be determined. In principle, it is
also possible to include the effect of an electromagnetic field. In
particular, we are going to consider a spherically symmetric electromagnetic
field, whose components are%
\begin{equation}
E_{r}=E_{1}\left( r\right) =cF_{01}=-cF_{10}.
\end{equation}%
All the other components are zero since we will not consider currents or magnetic
monopoles. This means that the electric field can only have a radial
component. Also, this radial component must not depend on $\theta $ or $\phi 
$. With this assumptions, in an orthonormal frame, we can write%
\begin{gather}
T_{\mu \nu }^{EM}=\frac{1}{\mu _{0}}\left( g_{\nu \alpha }F_{\mu \gamma
}F^{\alpha \gamma }-\frac{1}{4}g_{\mu \nu }F_{\alpha \beta }F^{\alpha \beta
}\right)  \notag \\
=\frac{Q^{2}}{2\left( 4\pi \right) ^{2}\varepsilon _{0}r^{4}}%
diag(1,-1,1,1)=diag(\rho _{E},p_{r,E},p_{t,E},p_{t,E}).  \label{TEM}
\end{gather}%
It is interesting to observe that the SET $T_{\mu \nu }^{EM}$ satisfies the
following property%
\begin{equation}
\rho _{E}+p_{r,E}=0.
\end{equation}

\section{Casimir Wormhole}

\label{p3}The original Casimir Wormhole is a solution of the semiclassical EFE\cite{CW}%
. The SET $\left( \ref{SET}\right) $ has been adopted with the following
replacement $a\rightarrow r$. In other words, the distance between the
plates instead of being considered parametrically fixed is considered as a
radial coordinate to be integrated. If we assume that an Equation of State
(EoS) $p_{r}\left( r\right) =\omega \rho _{C}\left( r\right) $ is imposed,
then we find the following solution to the semiclassical EFE%
\begin{align}
\Phi (r)& =\frac{1}{2}\left( {\omega -1}\right) {\ln }\left( {\frac{r\omega 
}{\left( \omega r+r_{0}\right) }}\right)  \\
b(r)& =\left( 1-\frac{1}{\omega }\right) r_{0}+\frac{r_{0}^{2}}{\omega r},
\end{align}%
where%
\begin{equation}
{\omega =}\frac{r_{0}^{2}}{r_{1}^{2}};\qquad r_{1}^{2}=\frac{\pi
^{3}l_{p}^{2}}{90}.
\end{equation}%
This specific choice avoids the presence of a horizon. In particular, when ${%
\omega =3}$, one finds%
\begin{align}
\Phi (r)& ={\ln }\left( {\frac{3r}{\left( 3r+r_{0}\right) }}\right) 
\label{PhiCW} \\
b(r)& =\frac{2}{3}r_{0}+\frac{r_{0}^{2}}{3r}.  \label{bCW}
\end{align}%
The EFE are consistent provided one introduces an additional EoS between $%
p_{t}\left( r\right) $ and $\rho _{C}\left( r\right) $. This additional EoS
must be inhomogeneous, namely%
\begin{equation}
p_{t}(r)=\omega _{t}\left( r\right) \rho _{C}\left( r\right) {,}
\end{equation}%
where we have introduced a inhomogeneous EoS on the transverse pressure with%
\begin{equation}
\omega _{t}\left( r\right) =-\frac{{\omega }^{2}\left( 4r-r_{0}\right)
+r_{0}\left( 4\omega +1\right) }{4\left( \omega r+r_{0}\right) }\underset{%
\omega =3}{=}-\frac{9\left( 4r-r_{0}\right) +r_{0}\left( 13\right) }{4\left(
3r+r_{0}\right) }\underset{\omega =3}{=}-\frac{9r+r_{0}}{3r+r_{0}}.
\label{otr}
\end{equation}%
However, instead of an inhomogeneous EoS, we can use the unknown temperature
tensor to find%
\begin{equation}
\tau _{\rho }=0;\qquad \tau _{r}=0;\qquad \tau _{t}\left( r\right) =\frac{%
2r_{0}^{2}}{\kappa r^{3}\left( 3r+r_{0}\right) }.  \label{TSET}
\end{equation}%
In this way, the complete SET is given by%
\begin{equation}
T_{\mu \nu }=\rho u_{\mu }u_{\nu }+p_{r}n_{\mu }n_{\nu }+\left( p_{t}+\tau
_{t}\right) \sigma _{\mu \nu }
\end{equation}%
which makes the EFE consistent. Note that the thermal stress-energy tensor actually is a thermal stress tensor because $\tau_{\rho}=0$

\section{Casimir Wormhole with an additional spherically symmetric
electromagnetic field}

\label{p4}In this section we consider the contribution of a spherically symmetric electromagnetic field, described by the SET $\left( \ref{TEM}\right)$ to the SET $\left( \ref{Tmn}\right) $\cite{CCW}. We find%
\begin{equation}
T_{\mu \nu }=\left( \rho +\tau _{\rho }+\rho _{E}\right) u_{\mu }u_{\nu
}+\left( p_{r}+\tau _{r}+p_{r,E}\right) n_{\mu }n_{\nu }+\left( p_{t}+\tau
_{t}+p_{t,E}\right) \sigma _{\mu \nu }.
\end{equation}%
Because of the structure of the SET $\left( \ref{TEM}\right) $, we can see
that%
\begin{equation}
\rho \left( r\right) +p_{r}\left( r\right) =\rho _{C}\left( r\right)
+p_{r,C}\left( r\right) +\rho _{E}\left( r\right) +p_{r,E}\left( r\right) =-%
\frac{4\hbar c\pi ^{2}}{720r^{4}}<0,
\end{equation}%
where%
\begin{eqnarray}
\rho _{C}\left( r\right)  &=&-\frac{\hbar c\pi ^{2}}{720r^{4}};\qquad
p_{r,C}\left( r\right) =-\frac{3\hbar c\pi ^{2}}{720r^{4}};  \notag \\
\rho _{E}\left( r\right)  &=&\frac{Q^{2}}{2\left( 4\pi \right)
^{2}\varepsilon _{0}r^{4}};\qquad p_{r,E}\left( r\right) =-\frac{Q^{2}}{%
2\left( 4\pi \right) ^{2}\varepsilon _{0}r^{4}}.
\end{eqnarray}%
This means that the NEC is always violated. In this context, the total
energy density is represented by%
\begin{equation}
\rho \left( r\right) =\rho _{C}\left( r\right) +\rho _{E}\left( r\right) =-%
\frac{\hbar c\pi ^{2}}{720r^{4}}+\frac{Q^{2}}{2\left( 4\pi \right)
^{2}\varepsilon _{0}r^{4}}=-\frac{r_{1}^{2}}{\kappa r^{4}}+\frac{r_{2}^{2}}{%
\kappa r^{4}},  \label{rho(r)}
\end{equation}%
where%
\begin{align}
r_{1}^{2}& =\frac{\pi ^{3}l_{p}^{2}}{90},  \label{r1} \\
r_{2}^{2}& =\frac{GQ^{2}}{4\pi c^{4}\varepsilon _{0}}.  \label{r2}
\end{align}%
The solution for the EFE $\left( \ref{rho}\right) $ and the EFE $\left( \ref%
{pr}\right) $ leads to%
\begin{equation}
b\left( r\right) =r_{0}-\frac{r_{1}^{2}-r_{2}^{2}}{r_{0}}+\frac{%
r_{1}^{2}-r_{2}^{2}}{r}  \label{b(r)0}
\end{equation}%
and%
\begin{equation}
\Phi \left( r\right) =\frac{r_{1}^{2}+r_{2}^{2}}{r_{1}^{2}-r_{2}^{2}}\ln
\!\left( \frac{r_{0}r}{r_{0}r+r_{1}^{2}-r_{2}^{2}}\right) .  \label{Phi(r)}
\end{equation}%
The integration constant for $\Phi \left( r\right) $ has been chosen in such
a way that for $r\rightarrow \infty $, $\Phi \left( r\right) \rightarrow 0$.
The relationship between the throat radius and the two lengths appearing
into the shape function and the redshift function is%
\begin{equation}
r_{0}=\sqrt{3r_{1}^{2}+r_{2}^{2}}.\label{r0EM}
\end{equation}%
Since $r_{2}$ is defined by Eq.$\left( \ref{r2}\right) $, it is
straightforward to see that $r_{0}$ can change the size when the charge $Q$
grows up. Like for the pure CW, the rest of the EFE can be solved if we
assume that the temperature tensor has the following structure%
\begin{equation}
\tau _{\rho }=0;\qquad \tau _{r}=0;\qquad \tau _{t}\left( r\right) =\frac{%
2r_{1}^{2}\left( r_{0}\,r-2r_{2}^{2}\right) }{\left(
r_{0}\,r+r_{1}^{2}-r_{2}^{2}\right) \kappa \,r^{4}}.  \label{TSETC}
\end{equation}%
It is immediate to see that for $r_{2}=0$, the temperature tensor $\left( %
\ref{TSETC}\right) $ is in agreement with the temperature tensor $\left( \ref%
{TSET}\right) $.

\section{Reexamining the original Casimir structure}

\label{p5}If one solves the EFE $\left( \ref{rho}\right) $ with the energy density
expressed by Eq. $\left( \ref{rhoC}\right) $, one finds\footnote{%
See Ref.\refcite{GABTW} for further details.}%
\begin{equation}
b\left( r\right) =r_{0}-\frac{r_{1}^{2}}{3a^{4}}\left(
r^{3}-r_{0}^{3}\right) ,  \label{b(r)C}
\end{equation}%
where the plates separation $a$ is considered as fixed. The shape function $%
\left( \ref{b(r)C}\right) $ does not represent a TW in a strict sense
because the asymptotic flatness is missing. However, we can also observe
that $b\left( r\right) $ described by Eq.$\left( \ref{b(r)C}\right) $ has
one real root. This is located at%
\begin{equation}
b\left( \bar{r}\right) =0\qquad \Longleftrightarrow \qquad \bar{r}=r_{0}\sqrt%
[3]{1+\frac{3a^{4}}{r_{0}^{2}r_{1}^{2}}}.
\end{equation}%
Since $b\left( r\right) <0$ is meaningless, we can assume the following
setting%
\begin{align}
b(r)& =r_{0}-\frac{r_{1}^{2}}{3a^{4}}\left( r^{3}-r_{0}^{3}\right) ,\qquad
\Phi (r)=0;\qquad r_{0}\leq r\leq \bar{r}  \notag \\
b(r)& =0,\qquad \Phi (r)=0;\qquad r\geq \bar{r}.  \label{CbP}
\end{align}%
Actually, it is not necessary to assume $\Phi (r)=0$. However, such a choice
gives the setting $\left( \ref{CbP}\right) $ a structure similar to an
Absurdly Benign Traversable Wormhole (ABTW) defined by%
\begin{align}
b(r)& =r_{0}\left( 1-\mu \left( r-r_{0}\right) \right) ^{2},\qquad \Phi
(r)=0;\qquad r_{0}\leq r\leq \bar{r}=r_{0}+\frac{1}{\mu }  \notag \\
b(r)& =0,\qquad \Phi (r)=0;\qquad r\geq \bar{r}.
\end{align}%
The similarity with the ABTW becomes stronger when we are close to the
throat since%
\begin{align}
b(r)& \simeq r_{0}\left( 1-\mu \left( r-r_{0}\right) \right) ,\qquad
r\rightarrow r_{0};\qquad r_{0}\leq r\leq \bar{r}  \notag \\
\mu & =\frac{r_{1}^{2}r_{0}}{a^{4}},\qquad \bar{r}=r_{0}+\frac{a^{4}}{%
r_{1}^{2}r_{0}}.  \label{ABTWa}
\end{align}%
It is straightforward to see that the ratio $a^{4}/r_{1}^{2}$ is huge. This
means that, to have consistency, $r_{0}$ must be huge too. Contrary to the
setting $\left( \ref{CbP}\right) $, the boundary of the approximated shape
function $\left( \ref{ABTWa}\right) $ allows to be determined. Indeed, if%
\begin{equation}
\bar{r}=r_{0}+\frac{a^{4}}{r_{1}^{2}r_{0}},
\end{equation}%
its minimum appears when%
\begin{equation}
r_{0,m}=\frac{a^{2}}{r_{1}}
\end{equation}%
and $\bar{r}$ becomes%
\begin{equation}
\bar{r}_{m}=2\frac{a^{2}}{r_{1}}.  \label{rm}
\end{equation}%
We can assume that the approximated value $\bar{r}_{m}$ be valid also for
the setting $\left( \ref{CbP}\right) $. The ABTW has another interesting
feature: for $r\geq \bar{r}$, the SET is Minkowski. To fix $\Phi (r)=0$
everywhere, it is necessary to introduce an inhomogeneous Equation of State
(EoS) of the form%
\begin{equation}
p_{r}\left( r\right) =\omega \left( r\right) \rho \left( r\right) .
\label{Inhom}
\end{equation}%
From Eqs.$\left( \ref{rho}\right) $ and $\left( \ref{pr}\right) $, by
imposing%
\begin{equation}
b\left( r\right) +\kappa p_{r}\left( r\right) r^{3}=0,  \label{constr}
\end{equation}%
we find%
\begin{equation}
\omega \left( r\right) =-\frac{b\left( r\right) }{b^{\prime }\left( r\right)
r}  \label{o(r)}
\end{equation}%
with $\Phi (r)=0$. Eq.$\left( \ref{o(r)}\right) $ can be solved to give%
\begin{equation}
b(r)=r_{0}\,\exp \left[ -\int_{r_{0}}^{r}\,\frac{dr^{\prime }}{\omega
(r^{\prime })r^{\prime }}\right] \,.  \label{form}
\end{equation}%
From the shape function $\left( \ref{form}\right) $, we can compute $\rho
\left( r\right) $%
\begin{equation}
\rho \left( r\right) =-\frac{r_{0}}{\kappa r^{3}\omega \left( r\right) }%
\,\exp \left[ -\int_{r_{0}}^{r}\,\frac{dr^{\prime }}{\omega (r^{\prime
})r^{\prime }}\right] ,  \label{rho1}
\end{equation}%
$p_{r}\left( r\right) $%
\begin{equation}
p_{r}\left( r\right) =-\frac{r_{0}}{\kappa r^{3}}\exp \left[
-\int_{r_{0}}^{r}\,\frac{dr^{\prime }}{\omega (r^{\prime })r^{\prime }}%
\right]   \label{pr1}
\end{equation}%
and $p_{t}\left( r\right) $%
\begin{equation}
p_{t}(r)=\frac{r_{0}}{2\kappa r^{3}}\left( \frac{1}{\omega \left( r\right) }%
+1\right) \exp \left[ -\int_{r_{0}}^{r}\,\frac{dr^{\prime }}{\omega
(r^{\prime })r^{\prime }}\right] .  \label{pt1}
\end{equation}%
If the relationship $\left( \ref{o(r)}\right) $ is satisfied, then we have a
zero-tidal-force wormhole (ZTF), a condition represented by the choice $\Phi
(r)=0$. This is also the same condition assumed for the ABTW. In particular,
for the ABTW, it is immediate to obtain that%
\begin{gather}
\omega \left( r\right) =\frac{1}{2r\mu }\left( 1-\mu \left( r-r_{0}\right)
\right) ,  \notag \\
\omega \left( r_{0}\right) =\frac{1}{2\mu r_{0}}\qquad \qquad \omega \left( 
\bar{r}\right) =0.
\end{gather}%
To complete the discussion we include also the form of the SET, written in
an orthonormal frame. We can write%
\begin{equation}
T_{\mu \nu }=\rho \left( r\right) diag\left( 1,\omega \left( r\right) ,-%
\frac{1}{2}-\frac{\omega \left( r\right) }{2},-\frac{1}{2}-\frac{\omega
\left( r\right) }{2}\right) .  \label{SETo}
\end{equation}%
By construction the SET $\left( \ref{SETo}\right) $ is divergenceless, but
it is not traceless. Nevertheless, this choice does not lead to a Minkowski
space outside the region located at $r=\bar{r}$, because $\rho \left(
r\right) $ does not vanish for  $r>\bar{r}$. However, it is immediate to
realize that the profile described by $\left( \ref{CbP}\right) $ can be
generalized to look like an ABTW\cite{GABTW}%
\begin{align}
b(r)& =\left[ r_{0}-\frac{r_{1}^{2}}{6a^{4}}\left( r^{3}-r_{0}^{3}\right) %
\right] ^{2};\qquad \Phi (r)=0;\qquad r_{0}\leq r\leq \bar{r}  \notag \\
b(r)& =0,\qquad \Phi (r)=0;\qquad r\geq \bar{r},  \label{GCbP}
\end{align}%
where, this time%
\begin{equation}
\bar{r}=r_{0}\sqrt[3]{1+\frac{6a^{4}}{r_{0}^{2}r_{1}^{2}}}.  \label{Sol}
\end{equation}%
In this way by imposing the EoS we find%
\begin{align}
\omega \left( r\right) & =\frac{2a^{4}}{r^{3}r_{1}^{2}}\left( r_{0}-\frac{%
r_{1}^{2}}{6a^{4}}\left( r^{3}-r_{0}^{3}\right) \right)   \notag \\
\omega \left( r\right) & =0\qquad r\geq \bar{r},\qquad \qquad \omega \left(
r_{0}\right) =\frac{2a^{4}}{r_{0}^{2}r_{1}^{2}}.
\end{align}%
The corresponding SET can be derived from the Eq.$\left( \ref{SETo}\right) 
$ where%
\begin{equation}
\rho \left( r\right) =-\rho _{0}\left[ r_{0}-\frac{r_{1}^{2}}{6a^{4}}\left(
r^{3}-r_{0}^{3}\right) \right] \qquad \Longrightarrow \qquad \rho \left(
r_{0}\right) =-\rho _{0}=-\frac{r_{1}^{2}}{\kappa a^{4}},
\end{equation}%
namely the Casimir energy density. Note that outside the region located at $%
r=\bar{r}$, the spacetime is Minkowski. Moreover by fixing%
\begin{equation}
\omega \left( r_{0}\right) =\frac{2a^{4}}{r_{0}^{2}r_{1}^{2}}=3,
\label{o(r)C}
\end{equation}%
we can recover the Casimir structure of the SET and putting numbers inside
the previous relationship, one finds%
\begin{equation}
r_{0}=\sqrt{\frac{1}{\rho _{0}\kappa }}=\sqrt{\frac{2}{3}}\frac{a^{2}}{r_{1}}%
,
\end{equation}%
and the external boundary is located at%
\begin{equation}
\bar{r}=r_{0}\sqrt[3]{10}.
\end{equation}%
where we have used the constraint $\left( \ref{o(r)C}\right) $.

\section{Casimir Wormhole at $T\neq 0$}

\label{p6}The energy density $\left( \ref{rhoC}\right) $ has been obtained
at $T=0$. If one is interested in correction induced by a temperature $T\neq
0$, we have to introduce a critical temperature defined by%
\begin{equation}
T_{e}=\frac{\hbar c}{2ak_{B}}.  \label{Teff}
\end{equation}%
To have an order of magnitude on $T_{e}$, we can use estimated value for
different quantities, and obtain%
\begin{equation}
T_{e}=\frac{\hbar c}{2k_{B}a}\simeq \frac{\left( 10^{-34}\mathrm{J\ s}%
\right) \left( 10^{8}m\mathrm{\ s}^{-1}\right) }{2a\left( 10^{-23}\mathrm{J\
K}^{-1}\right) }\simeq \frac{10^{-3}}{a}m\ K.
\end{equation}%
Therefore for plates separated by a distance of the order $10^{-6}m$, we
obtain $T_{e}\simeq 5\times 10^{2}K$. We can consider two possibilities:%
\footnote{%
See Ref.\cite{RGMF}for further details and also Refs.\cite{HotC,HotC1,HotC2}
for other frameworks.} Low-Temperature corrections to the Casimir energy and
High-Temperature corrections to the Casimir energy. The Low-Temperature
corrections to the Casimir energy do not modify significantly the results
obtained at $T=0$. This claim is valid when the plates are considered
parametrically fixed. Indeed, the case where the plates are radially varying
does not produce a shape function which is asymptotically flat. This small
modification is really important because it witnesses that the temperature
effects on the Casimir device do not destroy the property of supporting a
TW. On the other hand, for the High Temperature corrections we take into
account only the leading order corrections, since the other terms are
exponentially suppressed. For the pressure, one finds%
\begin{equation}
P\left( a,T\right) =-\frac{k_{B}T}{4\pi a^{3}}\zeta \left( 3\right) ,
\label{PH}
\end{equation}%
while for the energy, we get%
\begin{equation}
\qquad E\left( a,T\right) =-\frac{k_{B}TS}{8\pi a^{2}}\zeta \left( 3\right) ,
\end{equation}%
where $\zeta \left( x\right) $ is the Riemann zeta function. As in the
low-temperature approximation, we can consider two possible configurations:

\begin{enumerate}
\item We divide $E\left( a,T\right) $ with a volume term of the form $V=Sa$,
leading to the following form of the energy density%
\begin{equation}
\rho _{H,1}\left( a,T\right) =-\frac{k_{B}T}{8\pi a^{3}}\zeta \left(
3\right) .  \label{rhoH1}
\end{equation}%
The related pressure is represented by Eq.$\left( \ref{PH}\right) $ leading
to the following EoS%
\begin{equation}
\frac{P\left( a,T\right) }{\rho _{H,1}\left( a,T\right) }=\omega _{H}=2.
\label{omegaH}
\end{equation}%
Note the difference with respect to the $T=0$ case where $\omega =3$.

\item Using previous work on zero temperature Casimir wormholes \cite{CW},
we promote the distance separating plates $d$ to a radial variable $r$, and
we divide $E\left( r,T\right) $ with a volume term of the form $V=Sr$. Thus,
we obtain the energy density as%
\begin{equation}
\rho_{H,2}\left( r,T\right) =-\frac{k_{B}T}{8\pi r^{3}}\zeta\left( 3\right) .
\label{rho2}
\end{equation}
Even in this case, we have that the EoS leads to the same value of Eq.$%
\left( \ref{omegaH}\right) $, namely $\omega_{H}=2$.
\end{enumerate}

\subsection{Constant Plates Separation}

In this section, we consider the profile $\left( \ref{rhoH1}\right) $, which
produces a shape function of the form%
\begin{equation}
b\left( r\right) =r_{0}-\frac{8\pi G}{c^{4}}\left( \frac{k_{B}T}{8\pi a^{3}3}%
\right) \zeta \left( 3\right) \left( r^{3}-r_{0}^{3}\right) =r_{0}-\frac{%
l_{P}^{2}\zeta \left( 3\right) }{6a^{4}}\left( \frac{T}{T_{e}}\right) \left(
r^{3}-r_{0}^{3}\right) .  \label{b(r)HC}
\end{equation}%
It is easy to check that, even in this case, the flare-out condition is
always satisfied, 
\begin{equation}
b^{\prime }\left( r_{0}\right) =-\frac{l_{P}^{2}\zeta \left( 3\right) }{%
2a^{4}}\left( \frac{T}{T_{e}}\right) r_{0}^{2}<1.
\end{equation}%
Now by looking at its analytic form, we can observe that there exists an $%
\bar{r}$, such that $b\left( \bar{r}\right) =0$. Indeed, we obtain%
\begin{equation}
\bar{r}=r_{0}\sqrt[3]{1+\frac{l_{1}^{2}\left( a,T\right) }{r_{0}^{2}}}
\label{rbar}
\end{equation}%
where we have defined%
\begin{equation}
l_{1}\left( a,T\right) =\frac{a^{2}}{l_{P}}\sqrt{\frac{6}{\zeta \left(
3\right) }\left( \frac{T_{e}}{T}\right) }\simeq \frac{5\times 10^{24}}{\sqrt{%
T}}m  \label{l(a,T)}
\end{equation}%
Here we have considered distance between the plates of order $a\simeq
10^{-6}m$ and $T_{e}$ $\simeq 5\times 10^{2}\ ^{\circ }K$. Note that for
very high $T$, $l\left( a,T\right) \rightarrow 0$. Plugging Eq.$\left( \ref%
{b(r)HC}\right) $ into Eq.$\left( \ref{pr}\right) $, we obtain%
\begin{equation}
\Phi ^{\prime }\!\left( r\right) =\frac{r_{0}^{3}+r_{0}l_{1}^{2}\left(
a,T\right) -\left( 3\omega _{H}+1\right) r^{3}}{2r\left( r\left(
r^{2}+l_{1}^{2}\left( a,T\right) \right) -r_{0}\left( l_{1}^{2}\left(
a,T\right) +r_{0}^{2}\right) \right) }.  \label{Phi'(r)H}
\end{equation}%
Close to the throat, we obtain 
\begin{equation}
\Phi ^{\prime }\!\left( r\right) \simeq \frac{l_{1}^{2}\left( a,T\right)
-3\omega r_{0}^{2}}{2\left( l_{1}^{2}\left( a,T\right) +r_{0}^{2}\right)
\left( r-r_{0}\right) }.
\end{equation}%
Therefore, the horizon is not formed, if we choose%
\begin{equation}
\omega _{H}=\frac{l_{1}^{2}\left( a,T\right) }{3r_{0}^{2}}.  \label{oHH}
\end{equation}%
Moreover, from Eq.$\left( \ref{omegaH}\right) $, we can determine the size
of the throat. Indeed, we find%
\begin{equation}
r_{0}=\frac{\sqrt{6}l_{1}\left( a,T\right) }{6}\simeq \frac{5.6\times 10^{19}%
}{\sqrt{T}}m.
\end{equation}%
Thus, the effect of high-temperature corrections on the Casimir energy is a
reduction of the throat size. Nevertheless, we have to observe that in a
laboratory $T\simeq 10^{8}\ K$ as an ideal result and this implies that $%
r_{0}\simeq 10^{11}m$, which is an order of magnitude bigger than the solar
system size.

\subsection{Variable Plates separation}

The energy density $\left( \ref{rho2}\right) $ can be rewritten in the
following way%
\begin{equation}
\rho _{H,2}\left( r,T\right) =-\frac{\hbar c}{16\pi r^{3}}\frac{\zeta \left(
3\right) }{\lambda _{C}\left( T\right) }.
\end{equation}%
Here we have introduced the Casimir thermal wavelength%
\begin{equation}
\lambda _{C}\left( T\right) =\frac{\hbar c}{2k_{B}T}.
\end{equation}%
In the high-temperature approximation, or long-distance approximation, the
following inequality is satisfied%
\begin{equation}
\frac{2\pi }{\lambda _{C}\left( T\right) }a\gg 1,
\end{equation}%
where $d$ is the distance separating the plates. By promoting the $d$
distance to a variable distance $r$, we can write%
\begin{equation}
\frac{1}{\lambda _{C}\left( T\right) }=\frac{A}{2\pi r},\qquad A\gg 1.
\end{equation}%
Thus, $\rho _{H,2}\left( r,T\right) $ can be cast into the form 
\begin{equation}
\rho _{H,2}\left( r,T\right) =-A\frac{\hbar c}{32\pi ^{2}r^{4}}\zeta \left(
3\right) .
\end{equation}%
Thus the first EFE can be written as%
\begin{equation}
b\left( r\right) =r_{0}-\frac{\hbar G}{4\pi c^{3}}A\zeta \left( 3\right)
\int_{r_{0}}^{r}\frac{dr^{\prime }}{r^{\prime 2}}=r_{0}+r_{T}^{2}\left( 
\frac{1}{r}-\frac{1}{r_{0}}\right) ,
\end{equation}%
where we have defined%
\begin{equation}
r_{T}^{2}=\frac{A}{4\pi }\zeta \left( 3\right) l_{P}^{2}.
\end{equation}%
As we can see, we have the same formal expression found in previous work on
zero temperature Casimir wormholes \cite{CW}. Therefore, if we solve the
second EFE $\left( \ref{pr}\right) $, we get%
\begin{equation}
\Phi ^{\prime }\!\left( r\right) =\frac{\left( \left( -\omega +1\right)
r_{0}-r\right) r_{T}^{2}+rr_{0}^{2}}{2r\left( r-r_{0}\right) \left(
rr_{0}+r_{T}^{2}\right) }
\end{equation}%
and close to the throat we find that a horizon can be avoided if%
\begin{equation}
\omega =\frac{r_{0}^{2}}{r_{T}^{2}}.
\end{equation}%
However, this time the value of $\omega $ is given by%
\begin{equation}
\frac{P\left( r,T\right) }{\rho _{H,2}\left( r,T\right) }=\omega =2
\end{equation}%
and not $\omega =3$, as in previous work on zero temperature Casimir
wormholes \cite{CW}. So, we can now estimate the size of the throat as%
\begin{equation}
r_{0}=\sqrt{\frac{A}{2\pi }}l_{P}.
\end{equation}%
Compared to the zero temperature result \cite{CW}, we can observe that the
effect of the temperature is to enlarge the size of the throat.

\section{Casimir wormholes with an additional massless scalar field}

\label{p7}In this section we report a particular aspect discussed in Ref.\refcite{RGAZ},
where the Casimir source has been considered together a massless scalar
field. The equation of motion governing a massless scalar field simply is%
\begin{equation}
\nabla ^{2}\psi =\frac{1}{\sqrt{-g}}\partial _{\mu }\left( g^{\mu \nu }\sqrt{%
-g}\partial _{\nu }\right) \psi =0.
\end{equation}%
If we assume that $\psi $ is depending only on the radial coordinate $r$,
namely $\psi \equiv \psi \left( r\right) $, we can write%
\begin{equation}
\frac{\psi ^{\prime \prime }\left( r\right) }{\psi ^{\prime }\left( r\right) 
}+\frac{2}{r}+\Phi ^{\prime }(r)+\frac{\left( 1-\frac{b(r)}{r}\right)
^{\prime }}{2\left( 1-\frac{b(r)}{r}\right) }=0
\end{equation}%
or%
\begin{equation}
e^{2\Phi (r)}r^{4}\left( 1-\frac{b(r)}{r}\right) \psi ^{\prime }(r)^{2}=C\ (%
\mathrm{const.})\quad \mathrm{with}\quad C>0\,.  \label{eom_sc}
\end{equation}%
Note that the case with a negative value for the integration constant $C$ ($%
C<0$) corresponds to the introduction of a phantom field which we want to
avoid. Regarding the Casimir apparatus, we distinguish two different cases:

\begin{description}
\item[a)] the fixed Casimir plates case and

\item[b)] the variable Casimir plates case.
\end{description}

Here we will discuss the variable Casimir plates case and we shall refer the
reader to Ref.\refcite{RGAZ} for details. For the variable plates choice, the
total SET can be written as%
\begin{align}
\rho \left( r\right) & =\frac{1}{2}\left( 1-\frac{b(r)}{r}\right) \psi
^{\prime }(r)^{2}-\frac{\hbar c\pi ^{2}}{720r^{4}}+\tau _{\rho }\left(
r\right) =\frac{Ce^{-2\Phi (r)}}{2r^{4}}-\frac{r_{1}^{2}}{\kappa r^{4}}+\tau
_{\rho }\left( r\right)  \label{rhor} \\
p_{r}\left( r\right) & =\frac{1}{2}\left( 1-\frac{b(r)}{r}\right) \psi
^{\prime }(r)^{2}-\frac{3\hbar c\pi ^{2}}{720r^{4}}+\tau _{r}\left( r\right)
=\frac{Ce^{-2\Phi (r)}}{2r^{4}}-\frac{3r_{1}^{2}}{\kappa r^{4}}+\tau
_{r}\left( r\right)  \label{prr} \\
p_{t}\left( r\right) & =-\frac{1}{2}\left( 1-\frac{b(r)}{r}\right) \psi
^{\prime }(r)^{2}+\frac{\hbar c\pi ^{2}}{720r^{4}}+\tau _{t}\left( r\right)
=-\frac{Ce^{-2\Phi (r)}}{2r^{4}}+\frac{r_{1}^{2}}{\kappa r^{4}}+\tau
_{t}\left( r\right)  \label{ptr}
\end{align}%
upon using $\left( \ref{eom_sc}\right) $. Even in this case we will consider
the contribution of the thermal tensor introduced in section \ref{p2}. Since
we are mixing the effects of a positive scalar field and the Casimir source,
we impose at least one of the two features of a CW.

\subsection{Fixing the Casimir Wormhole redshift function}

If we assume the validity of the redshift function $\left( \ref{PhiCW}%
\right) $, then it is possible to determine the structure of the shape
function with the help of the first EFE. To this purpose, Eq.$\left( \ref%
{rho}\right) $ can be cast into the form%
\begin{equation}
\frac{b^{\prime }\left( r\right) }{r^{2}}=\frac{\kappa Ce^{-2\Phi (r)}}{%
2r^{4}}-\frac{r_{1}^{2}}{r^{4}}=\frac{\kappa C\left( 3r+r_{0}\right) ^{2}}{%
18r^{6}}-\frac{r_{1}^{2}}{r^{4}},  \label{rhoCWS}
\end{equation}%
where we have used Eq.$\left( \ref{rhor}\right) $ and where we have
considered $\tau _{\rho }\left( r\right) =0$. The energy density on the
r.h.s. of Eq.$\left( \ref{rhoCWS}\right) $, is such that%
\begin{equation}
\rho \left( r_{0}\right) \gtreqless 0\qquad \Longleftrightarrow \qquad
C\gtreqless \frac{3r_{0}^{2}}{8\kappa }.  \label{rho(r0)}
\end{equation}%
The integration of Eq.$\left( \ref{rhoCWS}\right) $ leads to%
\begin{equation}
b\!\left( r\right) =\frac{2r_{0}}{3}+\frac{r_{0}^{2}}{3r}-\frac{C\kappa }{2r}%
-\frac{\kappa Cr_{0}}{6r^{2}}-\frac{C\kappa r_{0}^{2}}{54r^{3}}+\frac{%
37C\kappa }{54r_{0}}.  \label{b(r)CWS}
\end{equation}%
The shape function in $\left( \ref{b(r)CWS}\right) $ has been obtained by
imposing $r_{0}=\sqrt{3}r_{1}$. It is immediate to recognize that, for $C=0$%
, one recovers the original CW. In order to satisfy the traversability, we
need to impose the flare-out condition, namely%
\begin{equation}
b\!^{\prime }\left( r_{0}\right) <1\qquad \Longleftrightarrow \qquad 0<C<%
\frac{3r_{0}^{2}}{2\kappa },
\end{equation}%
which is compatible with the relationship $\left( \ref{rho(r0)}\right) $.
Plugging $\left( \ref{b(r)CWS}\right) $ into the Eq.$\left( \ref{pr}\right) $%
, one finds%
\begin{equation}
\frac{2}{r}\left( 1-\frac{b\left( r\right) }{r}\right) \Phi \!^{\prime
}\left( r\right) -\frac{b\left( r\right) }{r^{3}}=\frac{\kappa C\left(
3r+r_{0}\right) ^{2}}{18r^{6}}-\frac{3r_{1}^{2}}{r^{4}}+\kappa \tau
_{r}\left( r\right) ,
\end{equation}%
leading to%
\begin{equation}
\tau _{r}\left( r\right) =-\frac{C\left(
37r^{3}+37r^{2}r_{0}-9rr_{0}^{2}-r_{0}^{3}\right) }{%
54r^{6}r_{0}+18r^{5}r_{0}^{2}}.
\end{equation}%
Finally, plugging $\left( \ref{b(r)CWS}\right) $ into the Eq.$\left( \ref{pt}%
\right) $, one finds%
\begin{eqnarray}
&&\left( 1-\frac{b\left( r\right) }{r}\right) \left[ \Phi ^{\prime \prime
}\!\left( r\right) +\Phi \!^{\prime }\left( r\right) \left( \Phi ^{\prime
}\!\left( r\right) +\frac{1}{r}\right) \right] -\frac{b^{\prime }\left(
r\right) r-b\left( r\right) }{2r^{2}}\left( \Phi \!^{\prime }\left( r\right)
+\frac{1}{r}\right)  \notag \\
&=&-\frac{\kappa Ce^{-2\Lambda (r)}}{2r^{4}}+\frac{r_{1}^{2}}{r^{4}}+\kappa
\tau _{t}\left( r\right) ,
\end{eqnarray}%
leading to%
\begin{equation}
\tau _{t}\!\left( r\right) =\frac{111C\kappa \,r^{4}+\left( 185\kappa
Cr_{0}+216r_{0}^{3}\right) r^{3}+\left( 72r_{0}^{4}-81C\kappa
r_{0}^{2}\right) r^{2}-21C\kappa rr_{0}^{3}-2C\kappa r_{0}^{4}}{%
36r^{5}\kappa r_{0}\left( 3r+r_{0}\right) ^{2}}.
\end{equation}%

\subsection{Fixing the Casimir Wormhole shape function}

In this other part of the section, we will consider the shape function of
the CW and we are going to determine the related redshift function which
solves the EFE. Thus, by plugging Eq.$\left( \ref{bCW}\right) $ into the
first EFE Eq.$\left( \ref{rho}\right) $, one finds%
\begin{equation}
-\frac{r_{0}^{2}}{3r^{4}}=\frac{\kappa Ce^{-2\Phi (r)}}{2r^{4}}-\frac{%
r_{1}^{2}}{r^{4}},
\end{equation}%
which implies%
\begin{equation}
\Phi (r)=\frac{1}{2}\ln \left( -\frac{3\kappa C}{2\left(
r_{0}^{2}-3r_{1}^{2}\right) }\right) .  \label{Phi(r)SCW}
\end{equation}%
Since $C>0,$we need to assume that%
\begin{equation}
3r_{1}^{2}>r_{0}^{2},
\end{equation}%
that it means that we cannot have a CW in the strict sense. However, we can
obtain additional interesting properties if we also assume that%
\begin{equation}
C=2\frac{3r_{1}^{2}-r_{0}^{2}}{3\kappa },
\end{equation}%
because this implies that $\Phi (r)=0$ without imposing any EoS. Plugging $%
\Phi (r)=0$ and Eq.$\left( \ref{bCW}\right) $ into the rest of the EFEs, one
gets%
\begin{equation}
\tau _{r}\left( r\right) =\frac{-3C\kappa -4rr_{0}-2r_{0}^{2}+18r_{1}^{2}}{%
6\kappa \,r^{4}}=\frac{2\left( 3r_{1}^{2}-rr_{0}\right) }{3\kappa \,r^{4}}
\end{equation}%
for the second EFE, Eq.$\left( \ref{pr}\right) $ and%
\begin{equation}
\tau _{t}\left( r\right) =\frac{r_{0}}{3\kappa r^{3}}
\end{equation}%
for the third EFE, Eq.$\left( \ref{pt}\right) $. Note that $\tau _{r}\left(
r\right) $ becomes negative when $r>r_{0}/\left( 3r_{1}^{2}\right) $ while $%
\tau _{t}\left( r\right) $ does not change its sign. The unpleasant feature
of this profile is that $0<r_{0}<\sqrt{3}r_{1}$ and there is no possibility
to have a larger value for $r_{0}$.

\section{Conclusions}
\label{p8}The idea of connecting Traversable Wormholes and the Casimir
energy has a long history\cite{Visser, MTY,FR}. However, the corresponding
space-time metric describing such a connection has been considered only
recently\cite{CW}. Since then, a certain degree of curiosity has been shown
about this research\cite%
{FW1,FW2,FW3,FW4,FW5,FW6,FW7,FW8,FW9,FW10,FW11,FW12,FW13,FW14}, especially
for the GUP modifications\cite{GUP,GUP1,GUP2,GUP3,GUP4,GUP5,GUP6,GUP7} and
modified gravity theories\cite%
{MG,MG1,MG2,MG3,MG4,MG5,MG6,MG7,MG8,MG9,MG10,MG11,MG12,MG13,MG14}.
Basically, a CW is a solution of the EFE obtained by imposing that the
source be represented by the SET $\left( \ref{SET}\right) $ leading to the
important relationship between the pressure $\left( \ref{P(a)}\right) $ and
the energy density $\left( \ref{rhoC}\right) $, namely 
\begin{equation}
\frac{P\left( a\right) }{\rho \left( a\right) }=3.
\end{equation}%
The second basic ingredient is the promotion of the plates separation $a$ to
a radial variable plates separation $r$. The result is described by the
shape function $\left( \ref{bCW}\right) $ and the redshift function $\left( %
\ref{PhiCW}\right) $. The consistency is guaranteed by introducing an
external stress tensor, which we have interpreted as a thermal stress tensor%
\cite{Hayward}. For a CW such a contribution is restricted only on $\tau
_{t}\left( r\right) $. The prediction of the throat size of such a TW is
Planckian. Things appear to be different when one includes the contribution
of a static electric field, because the throat size can change its value as
a function of the charge allowing, in principle, a TW with a very large
radius as showed by Eq.$\left( \ref{r0EM}\right) $. The introduction of a
temperature into the Casimir device to make the source more realistic does
not change very much the structure of the solutions, even if for the high
temperature case the EoS changes from $\omega =3$ to $\omega =2$. We do not
know why such a modification appears at this stage of the investigation. It
is important to observe that the corrections induced by a non-vanishing
temperature into a Casimir device do not destroy the traversability
properties. The same behavior seems to appear also for the combination of a
scalar field and a Casimir source as established in section \ref{p7} and in
Ref.\refcite{RGAZ}. A particular mention must be included about the Yukawa
deformation on a CW\cite{YC},where a deformation of the kind\footnote{%
see also Refs.\cite{YC1,YC2,YC3,YC4,YC5,YC6} for other approaches on
Yukawa-Casimir wormholes.}%
\begin{equation}
\rho \left( r\right) =\frac{r_{0}\rho _{C}}{r}\left( \alpha e^{-\mu \left(
r-r_{0}\right) }-\left( 1-\alpha \right) e^{-\nu \left( r-r_{0}\right)
}\right) \qquad \mu ,\nu >0.  \label{rhoab}
\end{equation}%
has been considered. Such a deformation on the energy density produces a
shape function of the form 
\begin{equation}
b\left( r\right) =r_{0}{\frac{\left( e{^{-\nu \left( r-r_{0}\right) }}%
A\left( \nu r+1\right) -e{^{-\mu \left( r-r_{0}\right) }B}\left( \mu
r+1\right) \right) }{\left( \nu r_{0}+1\right) {\mu }^{2}+\mu {\nu }%
^{2}r_{0}+{\nu }^{2}},}
\end{equation}%
which is useful to compute the function $\omega \left( r\right) $%
\begin{equation}
\omega \left( r\right) ={\frac{Ae{^{-\nu \left( r-r_{0}\right) }}\left( \nu
r+1\right) -Be{^{-\mu \left( r-r_{0}\right) }}\left( \mu r+1\right) }{{r}%
^{2}\left( {\nu }^{2}Ae{^{-\nu \left( r-r_{0}\right) }}-{\mu }^{2}Be{^{-\mu
\left( r-r_{0}\right) }}\right) },}
\end{equation}%
where we have defined%
\begin{eqnarray}
A &=&\rho \kappa \left( 1+\mu r_{0}\right) +{\mu }^{2},  \notag \\
B &=&\rho \kappa \left( 1+\nu r_{0}\right) -{\nu }^{2}.
\end{eqnarray}%
$\omega \left( r\right) $ is related to a EoS allowing to set the redshift
function $\Phi \left( r\right) =0$. Such a profile allows to have a TW with
a throat that can be fine tuned with respect to the original Casimir source.


\begin{thebibliography}{99}

\bibitem{Casimir} H. Casimir, \textsl{Proc. Kon. Ned. Akad. Wetenschap},
vol. \textbf{51}, p. 793, (1948). 

\bibitem{Sparnaay} M. Sparnaay, \textsl{Nature} \textbf{180}, p. 334,
(1957). 
M. Sparnaay, \textsl{Physica} \textbf{24}, p. 751, (1958). 

\bibitem{AVS} A. van Silfhout, \textsl{Drukkerij Holland}, N. V. ,
Amsterdam, (1966). 

\bibitem{Lamoreaux} S. Lamoreaux, \textsl{Phys. Rev. Lett.}, \textbf{78}, 5,
(1997). 

\bibitem{BCOR} G. Bressi, G. Carugno, R. Onofrio and G. Ruoso \textsl{Phys.
Rev. Lett.} \textbf{88}, 041804 (2002) 041804; arXiv:0203002 [quant-ph]. 

\bibitem{Boyer} T. Boyer, \textsl{Phys. Rev.} \textbf{174}, 1764 (1968). 

\bibitem{BVW} S. K. Blau, M. Visser, and A. Wipf, \textsl{Nucl. Phys.} 
\textbf{B}, \textbf{310}, 163, (1988). 

\bibitem{Hochberg} D. Hochberg, \textsl{Phys. Lett.} \textbf{B 268}, 377
(1991). 

\bibitem{Hayward} S. A. Hayward, \textit{Relativistic thermodynamics: From
Black Holes: New Horizons, pp. 175-201}, 2013.

\bibitem{MT} M.~S.~Morris and K.~S.~Thorne, \textsl{Am.\ J.\ Phys.}\ \textbf{%
56}, 395 (1988). 

\bibitem{Visser} M. Visser, \textit{Lorentzian Wormholes: From Einstein to
Hawking} (American Institute of Physics, New York), 1995.

\bibitem{MTY} M.S. Morris, K.S. Thorne and U. Yurtsever, \textsl{Phys. Rev.
Lett.} \textbf{61}, 1446 (1988). 

\bibitem{FR}  L.H. Ford and T. A. Roman, \textsl{Phys .Rev.} \textbf{D 53},
5496 (1996); arXiv:gr-qc/9510071. 

\bibitem{CW} R.Garattini, \textsl{Eur. Phys. J.} \textbf{C 79} (2019) 11,
951; arXiv: 1907.03623 [gr-qc]\textit{.} 

\bibitem{CCW} R. Garattini, \textsl{Eur. Phys. J.} \textbf{C 83}, 5, 369
(2023); arXiv: 2302.04043 [gr-qc].%

\bibitem{GABTW} R.~Garattini, \textsl{Eur. Phys. J.} \textbf{C 80}, 12, 1172
(2020); arXiv: 2008.05901 [gr-qc].%

\bibitem{RGMF} R. Garattini and M. Faizal, \textit{Hot Casimir Wormholes},
arXiv: 2403.15174 [gr-qc].

\bibitem{HotC} H. F.S. Mota, C. R. Muniz and V. B. Bezerra, \textsl{Universe}
8 (2022) 11, 597; arXiv: 2210.06128 [hep-th].%

\bibitem{HotC1} P. Channuie, \textsl{Nucl. Phys. }\textbf{B 1004} (2024)
116572; arXiv: 2403.19261 [hep-th].%

\bibitem{HotC2} V. B. Bezerra, H. F. Santana Mota, A. P.C.M. Lima, G.
Alencar and C. Rodrigues Muniz, \textsl{Physics} 6 (2024) 3, 1046-1071.

\bibitem{RGAZ} R. Garattini and A.G. Tzikas, \textquotedblleft \textit{%
Traversable Wormholes induced by Stress Energy Conservation: combining
Casimir Energy with a scalar field}\textquotedblright , arXiv: 2312.16736
[gr-qc].

\bibitem{FW1} W. Javed, A. Hamza and A. \"{O}vg\"{u}n. \textsl{Mod. Phys.
Lett.} \textbf{A 35} (2020) 39, 2050322; arXiv: 2101.04515 [gr-qc].%

\bibitem{FW2} S.K. Tripathy, \textsl{Phys. Dark Univ.} \textbf{31} (2021)
100757; arXiv: 2004.14801 [gr-qc].%

\bibitem{FW3} H. Huang and J. Yang,\textsl{\ Phys. Rev. }\textbf{D 100}
(2019) 12, 124063; arXiv: 1909.04603 [gr-qc].%

\bibitem{FW4} J. Matyjasek, \textsl{Phys. Rev. }\textbf{D 102} (2020) 2,
024082; arXiv: 2005.07077 [gr-qc].%

\bibitem{FW5} A.C.L. Santos, C.R. Muniz and L.T. Oliveira, \textsl{Int. J.
Mod. Phys. }\textbf{D 30} (2021) 05, 2150032; arXiv: 2007.00227 [gr-qc].%

\bibitem{FW6} G. Alencar, V.B. Bezerra(Paraiba U.) and C.R. Muniz, \textsl{%
Eur. Phys. J. }\textbf{C 81} (2021) 10, 924; arXiv: 2104.13952 [gr-qc].%
%

\bibitem{FW7} \'{I}. D.D. Carvalho, G. Alencar and  C.R. Muniz, \textsl{Int.
J. Mod. Phys. }\textbf{D 31} (2022) 03, 2250011; arXiv: 2106.11801 [gr-qc].%

\bibitem{FW8} P.H. F. Oliveira, G. Alencar, I.C. Jardim and R.R. Landim, 
\textsl{Mod. Phys. Lett. }\textbf{A 37} (2022) 15, 2250090; arXiv:
2107.00605 [hep-th].

\bibitem{FW9} R. Avalos, E. Fuenmayor and E. Contreras, \textsl{Eur. Phys.
J. }\textbf{C 82} (2022) 5, 420; arXiv: 2208.10261 [gr-qc].%

\bibitem{FW10} Z. Hassan, S. Ghosh, P.K. Sahoo and K. Bamba, \textsl{Eur.
Phys. J. }\textbf{C 82} (2022) 12, 1116; arXiv: 2207.09945 [gr-qc].%

\bibitem{FW11} C.R. Muniz, H.R. Christiansen, M.S. Cunha, J. Furtado and
V.B. Bezerra; \textsl{Universe} \textbf{8} (2022) 12, 616; arXiv: 2210.07427
[gr-qc].

\bibitem{FW12} R. Avalos and E. Contreras, \textsl{Annals Phys.} \textbf{446}
(2022) 169128; arXiv: 2302.09141 [gr-qc].%

\bibitem{FW13} A.C.L. Santos, R.V. Maluf and C.R. Muniz, \textsl{Annals Phys.%
} \textbf{469} (2024) 169775; arXiv: 2405.08774 [gr-qc].%

\bibitem{FW14} M. R. Mehdizadeh and A. H. Ziaie, \textquotedblleft \textit{%
Novel Casimir wormholes in Einstein gravity}\textquotedblright , arXiv:
2406.03588 [gr-qc].

\bibitem{GUP} K. Jusufi, P. Channuie and M. Jamil, \textsl{Eur. Phys. J.} 
\textbf{C 80} (2020) 2, 127; arXiv: 2002.01341 [gr-qc].%

\bibitem{GUP1} D. Samart, T. Tangphati and P. Channuie, \textsl{Nucl. Phys. }%
\textbf{B 980} (2022) 115848; arXiv: 2107.11375 [gr-qc].%

\bibitem{GUP2} S. Rani, M. B. A. Sulehri, A. Jawad and U. Zafar, \textsl{%
Int. J. Geom. Meth. Mod. Phys.} \textbf{20} (2022) 03.%

\bibitem{GUP3} A. Sahoo, S.K. Tripathy, B. Mishra and S. Ray, \textsl{Eur.
Phys. J. }\textbf{C 84} (2024) 3, 325; arXiv: 2308.06941 [gr-qc].%

\bibitem{GUP4} C. R. Muniz, T. Tangphati, R.M.P. Neves and M.B. Cruz, 
\textsl{Phys. Dark Univ.} \textbf{46} (2024) 101673; arXiv: 2406.08250
[gr-qc].%

\bibitem{GUP5} C. C. Chalavadi, V. Venkatesha and A. Malik, \textsl{Nucl.
Phys. }\textbf{B 1006} (2024) 116644.%

\bibitem{GUP6} M. M. Rizwan, Z. Hassan and P.K. Sahoo, \textquotedblleft 
\textit{GUP corrected Casimir Wormholes with Electric Charge in }$f\left(
R,L_{m}\right) $\textit{\ Gravity }\textquotedblright , arXiv: 2408.03969
[gr-qc].

\bibitem{GUP7} M. M. Rizwan, Z. Hassan, P.K. Sahoo and A. \"{O}vg\"{u}n,
\textquotedblleft \textit{Influence of GUP corrected Casimir energy on zero
tidal force wormholes in modified teleparallel gravity with matter coupling}%
\textquotedblright , arXiv: 2410.05348 [gr-qc].

\bibitem{MG} T. Tangphati, A. Chatrabhuti, D. Samart and P. Channuie, 
\textsl{Phys. Rev. }\textbf{D 102} (2020) 8, 084026; arXiv: 2003.01544
[gr-qc].

\bibitem{MG1} S. K. Tripathy, \textsl{Phys. Dark Univ.}\textbf{\ 31} (2021)
100757; arXiv: 2004.14801 [gr-qc].%

\bibitem{MG2} B. Mishra, A.S. Agrawal, S.K. Tripathy and S. Ray, \textsl{%
Int. J. Mod. Phys. }\textbf{D 30} (2021) 08, 2150061; arXiv: 2104.05440
[gr-qc].
%

\bibitem{MG3} V. De Falco, E. Battista, S. Capozziello and M. De Laurentis, 
\textsl{Phys. Rev. }\textbf{D 103} (2021) 4, 044007; arXiv: 2101.04960
[gr-qc].%

\bibitem{MG4} V. De Falco, E. Battista, S. Capozziello and M. De Laurentis, 
\textsl{Eur. Phys. J. }\textbf{C 81} (2021) 2, 157; arXiv: 2102.01123
[gr-qc].%

\bibitem{MG5} B. Mishra, A.S. Agrawal, S.K. Tripathy and S. Ray, \textsl{%
Int. J. Mod. Phys. }\textbf{A 37} (2022) 05, 2250010; arXiv: 2112.08365
[gr-qc].

\bibitem{MG6} O. Sokoliuk, A. Baransky and P.K. Sahoo, \textsl{Nucl. Phys. }%
\textbf{B 980} (2022) 115845; arXiv: 2205.14011 [physics.gen-ph].%

\bibitem{MG7} G. Mustafa, S.K. Maurya and Saibal Ray, \textsl{Astrophys. J.} 
\textbf{941} (2022) 2, 170.%

\bibitem{MG8} Z. Hassan, S. Ghosh, P.K. Sahoo and V. Sree Hari Rao, \textsl{%
Gen. Rel. Grav.} \textbf{55} (2023) 8, 90; arXiv: 2209.02704 [gr-qc].%

\bibitem{MG9} M. Zubair, S. Waheed, M. Farooq, A. H. Alkhaldi and A. Ali, 
\textsl{Eur. Phys. J. Plus} \textbf{138} (2023) 10, 902.%

\bibitem{MG10} A. Errehymy, \textsl{Phys. Dark Univ.} \textbf{44} (2024)
101438. 

\bibitem{MG11} M. Khatri and J. Lalvohbika, \textsl{Chin. J. Phys.} \textbf{%
89} (2024) 1222-1235.

\bibitem{MG12} A. Banerjee, S. Hansraj, A. Pradhan and A. Errehymy,
\textquotedblleft \textit{Is dark energy necessary for the sustainability of
traversable wormholes?}\textquotedblright\ arXiv: 2402.11348 [gr-qc].

\bibitem{MG13} A. H. Ziaie and M. R. Mehdizadeh, \textsl{Class. Quant. Grav.}
\textbf{41} (2024) 14, 145001; arXiv: 2406.10821 [gr-qc].%

\bibitem{MG14} A.S. Agrawal, S. Tarai, B. Mishra and S.K. Tripathy,
\textquotedblleft \textit{Matter Geometry Coupling and Casimir Wormhole
Geometry}\textquotedblright , arXiv: 2409.12160 [gr-qc].

\bibitem{YC} R. Garattini, \textsl{Eur. Phys. J. }\textbf{C 81} (2021) 9,
824; arXiv: 2107.09276 [gr-qc].

\bibitem{YC1} A. Jawad, M.Bilal Amin Sulehri and S. Rani, \textsl{Eur. Phys.
J. Plus} \textbf{137} (2022) 11, 1274.%

\bibitem{YC2} P. H. Ferreira de Oliveira, G. Alencar, I. Carneiro Jardim and
R. Renan Landim, \textsl{Symmetry} \textbf{15} (2023) 2, 383; arXiv:
2212.14664 [gr-qc].

\bibitem{YC3} A. K. Mishra, S. K. Sharma and U. K. Sharma, \textsl{Universe} 
\textbf{9} (2023) 4, 161; arXiv: 2303.04641 [physics.gen-ph].%

\bibitem{YC4} S. K. Mishra, U. K. Sharma and A. K. Mishra,\textsl{\ Int. J.
Geom. Meth. Mod. Phys.} \textbf{20} (2023) 08, 2350140.%

\bibitem{YC5} S. K. Mishra, U. K. Sharma and A. K. Mishra, \textsl{Int. J.
Geom. Meth. Mod. Phys. }\textbf{20} (2023) 13, 2350223.%

\bibitem{YC6} S. Waheed and M. Zubair, \textsl{Chin. J. Phys.} \textbf{89}
(2024) 1080-1101.%

\bibitem{YC7} V. Venkatesha, C. C. Chalavadi and A. Malik,\textsl{\ Eur.
Phys. J. }\textbf{C 84} (2024) 8, 834.%

\end{thebibliography}
\end{document}